\documentclass[twocolumn,aps,prx,superscriptaddress]{revtex4-2}
\usepackage[T1]{fontenc}
\usepackage[latin9]{inputenc}
\usepackage{textcomp}
\usepackage{bbding}
\usepackage{amsmath}
\usepackage{amssymb}
\usepackage{graphicx}
\usepackage{esint}
\usepackage{xcolor}
\usepackage[pdftex,colorlinks,linkcolor=blue,citecolor=blue,urlcolor=blue,filecolor=blue,breaklinks=true]{hyperref}

\makeatletter
\@ifundefined{textcolor}{}
{%
 \definecolor{BLACK}{gray}{0}
 \definecolor{WHITE}{gray}{1}
 \definecolor{RED}{rgb}{1,0,0}
 \definecolor{GREEN}{rgb}{0,1,0}
 \definecolor{BLUE}{rgb}{0,0,1}
 \definecolor{CYAN}{cmyk}{1,0,0,0}
 \definecolor{MAGENTA}{cmyk}{0,1,0,0}
 \definecolor{YELLOW}{cmyk}{0,0,1,0}
}

\makeatother

\begin{document}
\title{Practical Few-Atom Quantum Reservoir Computing}
\author{Chuanzhou Zhu}
\email{chuanzhouzhu@arizona.edu}
\affiliation{Wyant College of Optical Sciences, University of Arizona, Tucson, Arizona, USA}
\author{Peter J. Ehlers}
\affiliation{Wyant College of Optical Sciences, University of Arizona, Tucson, Arizona, USA}
\author{Hendra I. Nurdin}
\affiliation{School of Electrical Engineering and Telecommunications, University of New South Wales, Sydney, Australia}
\author{Daniel Soh}
\email{danielsoh@arizona.edu}
\affiliation{Wyant College of Optical Sciences, University of Arizona, Tucson, Arizona, USA}

\date{\today}
\begin{abstract}
Quantum Reservoir Computing (QRC) harnesses quantum systems to tackle intricate computational problems with exceptional efficiency and minimized energy usage. This paper presents a QRC framework that utilizes a minimalistic quantum reservoir, consisting of only a few two-level atoms within an optical cavity. The system is inherently scalable, as newly added atoms automatically couple with the existing ones through the shared cavity field. We demonstrate that the quantum reservoir outperforms traditional classical reservoir computing in both memory retention and nonlinear data processing through two tasks, namely the prediction of time-series data using the Mackey-Glass task and the classification of sine-square waveforms. Our results show significant performance improvements with an increasing number of atoms, facilitated by non-destructive, continuous quantum measurements and polynomial regression techniques. These findings confirm the potential of QRC as a practical and efficient solution to addressing complex computational challenges in quantum machine learning.
\end{abstract}
\maketitle

\section{Introduction}

As a transformative machine learning framework for the processing of temporal information, quantum reservoir computing (QRC) is utilized to tackle complex time-series tasks with outstanding efficiency and minimal energy consumption \cite{Fujii2017,chen2019learning,CNY20,Dudas2023,Govia2021,Bravo2022,Hulser2023,Nokkala2021,Pena2021,Xia2022,Mujal2023,Fry2023,Kalfus2022,Yasuda23,Beni2023,Markovic2019,Lin2020}. There are also static variants of QRC that process vectors or single quantum states, rather than time series data, which are referred to as quantum random kitchen sinks or quantum extreme learning machines \cite{Innocenti23}, as exemplified in \cite{Ghosh2019, Ghosh2021, Domingo2022}. The tasks addressed by QRC can be broadly categorized into two main types: classical tasks that necessitate quantum memories for the processing of nonlinear functions and datasets \cite{Fujii2017, Dudas2023, Govia2021}, and quantum tasks focused on recognizing quantum entanglements \cite{Ghosh2019}, measuring dispersive currents \cite{Angelatos2021}, and predicting molecular structures \cite{Domingo2022}. Among these categories, classical tasks offer a more straightforward basis for comparing the performance of QRC with that of classical reservoir computing (CRC) \cite{Tanaka2019, Appeltant2011, Pathak2018, Angelatos2021, Chen2022, Gauthier2021}. 

Various schemes for developing quantum reservoirs have been advanced across diverse platforms, including coupled qubit networks \cite{Fujii2017, chen2019learning, CNY20, Domingo2022, Ghosh2021, Pena2021, Xia2022, Yasuda23}, fermions \cite{Ghosh2019}, harmonic oscillators \cite{Nokkala2021}, Kerr nonlinear oscillators \cite{Angelatos2021}, Rydberg atoms \cite{Bravo2022}, and optical pulses \cite{Beni2023}. Increasing the number of physical sites, such as qubits, oscillators, or atoms, within a quantum network can result in a dramatic expansion of the Hilbert space, as the number of quantum basis states grows exponentially. Each of these quantum basis states functions as a node within the quantum neural network, analogous to the role of nodes in classical neural networks \cite{Fujii2017} or optical neural networks \cite{Wang2022, ma2023}. The exponential scalability of the quantum basis states is a fundamental advantage of QRC over CRC, with quantum phase transitions further enhancing computational performance \cite{Pena2021, Xia2022}. However, the proposed implementations face significant scalability challenges, primarily due to the inherent difficulty in establishing connections between newly introduced sites and the existing sites within a quantum network \cite{Fujii2017, chen2019learning, CNY20, Domingo2022, Ghosh2021, Pena2021, Ghosh2019, Xia2022, Yasuda23}. For example, to introduce a new qubit into the transverse-field Ising model, its coupling to the existing qubits needs to be engineered by randomly sampling from either a uniform distribution within a specified interval \cite{Fujii2017, Pena2021} or a Gaussian distribution \cite{Domingo2022}. In a recent QRC experiment using a Rydberg atom network, the interactions between Rydberg atoms had to be engineered by precisely controlling their spatial positions \cite{kornjaca2024}. Compared to these previous qubit-connection configurations, our scheme significantly reduces the increase in hardware complexity, as each newly added qubit (atom) naturally connects to all existing atoms through the shared cavity field.

Furthermore, several methods have been proposed for extracting information from quantum reservoirs, including the measurement of probabilities on quantum basis states \cite{Dudas2023}, excitations of qubits \cite{Fujii2017}, occupations of lattice sites \cite{Ghosh2019}, and energies of oscillators \cite{Angelatos2021}. Despite their applicability, these conventional measurement techniques present substantial obstacles to efficient data processing. These methods often require quantum tomography, a process that necessitates the complete destruction of the quantum reservoir to obtain features at each time step. To overcome this time complexity issue, several online protocols for QRC have been proposed. These include: online measurement based on partial readout and subsequent resetting of readout qubits to preserve memory of past inputs \cite{CNY20,Yasuda23,Hu2024}; weak and projective measurements to extract information accurately without hindering memory \cite{Mujal2023}; a feedback-driven QRC framework that feeds measurement outcomes back into the reservoir to restore memory of prior inputs \cite{Kobayashi2024}; and the application of singular value decomposition and data-filtering techniques to optimize the training of the finite-sampled QRC framework \cite{Ahmed2025}.

We introduce a quantum reservoir that offers both convenient scalability and practical measurability, achieved through the coupling of atoms and photons in an optical cavity. To assess its computing prowess, we evaluate its performance on two classical tasks: the Mackey-Glass task \cite{Fujii2017}, which demands long-term memory for predicting the future trend of an input function, and the sine-square waveform classification task \cite{Dudas2023,Markovic2019}, which requires nonlinearity of the reservoir to capture sudden, high-frequency shifts in a linearly inseparable input dataset. These classical inputs are embedded in the coherent driving of the cavity photon field, enabling the performance comparison between QRC and CRC. 

Our quantum reservoir is more conveniently scalable compared to prior proposals, since newly added atoms automatically couple with the existing atoms through their mutual connections with the cavity photon field. The performance improvement associated with this scalability is evidenced by the diminishing gap between the actual and target outputs as the number of atoms increases. The exponential scaling of basis states underlying the limited number of observable features plays an important role of making the system dynamics more nonlinear and more complex, ensuring the excellent performance of QRC.

Practical readouts are acquired through continuous quantum measurement, in which the cavity field generates two features corresponding to two photonic quadratures, while each atom yields two features associated with two atomic spin channels \cite{Wiseman2010,BvHJ07,Wei2008,Ruskov2010,Hacohen2016,Ochoa2018,Fuchs2001}. In contrast to the prior online protocols that require memory restoration, our continuous measurement scheme is considered non-destructive while its back-action on the reservoir is fully incorporated into our model, providing a convenient and practical method for information extraction from the reservoir. A continuous heterodyne measurement framework has been proposed for QRC to process quantum inputs as states of quantum systems \cite{khan2021}. We extend this continuous measurement approach to handle inputs as classical time-series functions.

\section{Setup}

\begin{figure}
\includegraphics[width=1.0\linewidth]{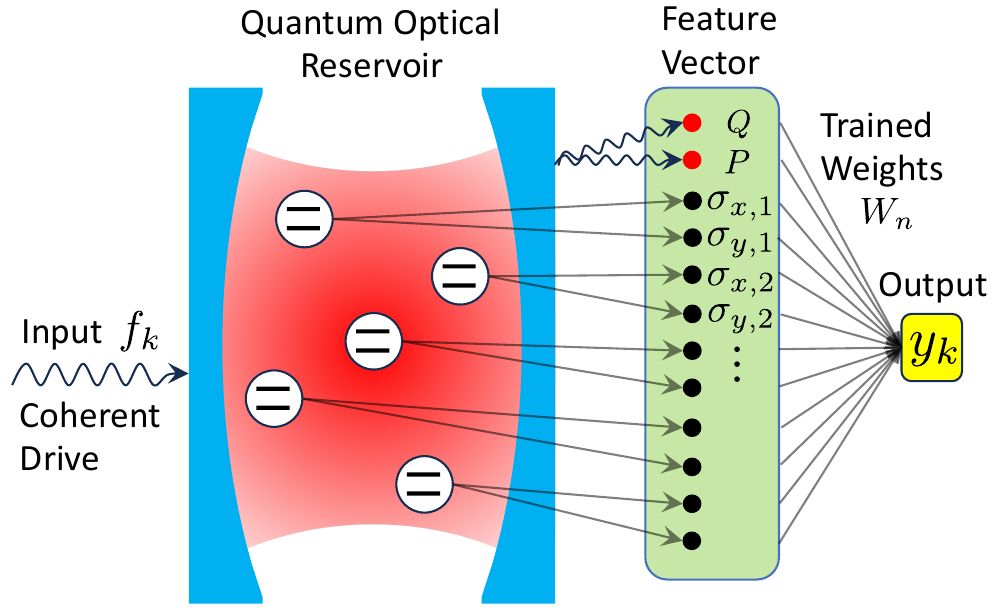}
\caption{Setup of quantum optical reservoir computing. The reservoir is composed of atoms inside an optical cavity, exhibiting diverse detunings and couplings across various spatial positions. The input function is integrated into the coherent driving of the cavity. Feature vector is obtained via continuous measurements of photonic quadratures and atomic spin channels. A machine learning process is employed to train the mapping from the feature vector to the output.}
\label{Scheme}
\end{figure}

The components of quantum reservoir computing include the input, the quantum reservoir, the feature vector, and the output, as illustrated in Fig.~\ref{Scheme}. We utilize a quantum optical system consisting of a few two-level atoms within a single-mode optical cavity, which serves as the quantum reservoir. The time-independent part of the Hamiltonian is given by 
\begin{equation}
H_{0}=\omega_{c}c^{\dagger}c+\underset{i}{\sum}\omega_{i}\sigma_{i}^{\dagger}\sigma_{i}+\underset{i}{\sum}g_{i}\left(c^{\dagger}\sigma_{i}+c\sigma_{i}^{\dagger}\right),\label{eq:H0}
\end{equation}
where $c$ denotes the photon annihilation operator, $\sigma_{i}=\left|g\right\rangle \left\langle e\right|_{i}$ represents the lowering operator for the $i$-th atom, with $\left|g\right\rangle $ ($\left|e\right\rangle $) representing the ground (excited) state, $\omega_{c}$ ($\omega_{i}$) characterizes the detuning between the coherent driving field and the cavity (atomic) frequencies, and $g_{i}$ is the electric-dipole coupling strength between the $i$-th atom and the cavity mode. In the context of QRC, it is crucial to vary either the detuning $\omega_{i}$ or the coupling strength $g_{i}$ in order to induce non-identical memory states in the atoms, thereby enhancing the overall performance of the system. This model can be experimentally realized using cold atoms in an optical cavity \cite{Brennecke2007, Niemczyk2010}, where optical tweezers can be employed to trap and measure individual atoms at distinct positions \cite{Kaufman2021, Ye2023}. Alternatively, quantum dots can be used to induce random variations in positioning, detuning, and coupling \cite{Schmidt2007}. The input function, denoted as $f\left(t\right)$, is integrated into the time-dependent coherent driving term
\begin{equation}
H_{1}\left(t\right)=i\epsilon f\left(t\right)\left(c-c^{\dagger}\right),\label{eq:H1}
\end{equation}
where $\epsilon$ is the driving strength. 

The features from the quantum optical reservoir, denoted as $x_{n}\left(t\right)$, are directly correlated with the experimental observables determined through continuous quantum measurement. The observables of the cavity field arise from the homodyne detection of two orthogonal quadratures \cite{Wiseman2010,BvHJ07,Nurdin14}
\begin{equation}
Q=c+c^{\dagger}, \quad P=i\left(c-c^{\dagger}\right); \label{eq:QP}
\end{equation}
and the observables of the atomic spontaneous emission are related to the Pauli operators \cite{Wiseman2001}
\begin{equation}
\sigma_{x,i}=\sigma_{i}+\sigma_{i}^{\dagger}, \quad \sigma_{y,i}=i\left(\sigma_{i}-\sigma_{i}^{\dagger}\right). \label{eq:sigmaxy}
\end{equation}
It has been shown that the observables of both the cavity and the atoms can be measured simultaneously  \cite{Wei2008,Ruskov2010,Hacohen2016,Ochoa2018}. 

For the cavity field, homodyne detection of two orthogonal quadratures is implemented by splitting the system's output beam using a beam splitter, followed by the homodyning of each beam with the same local oscillator, with a phase shift of $\pi/2$ between them \cite{Wiseman2010}. Similarly, for the atoms, homodyne detection is applied to the spontaneous emissions \cite{Wiseman2001}. These measurements can be performed simultaneously \cite{Wei2008,Ruskov2010,Hacohen2016,Ochoa2018}. As a result, the continuous measurement process is governed by the stochastic master equation \cite{VPB79,VPB88,VPB91a,HC93,Wiseman2010,BvHJ07,Nurdin14}.
\begin{align}
d\rho_{J} & =-i\left[H_{0}+H_{1}\left(t\right),\rho_{J}\right]dt\label{eq:sme}\\
 & +2\mathcal{D}\left[\sqrt{\kappa_{c}}c\right]\rho_{J}dt+2\underset{i}{\sum}\mathcal{D}\left[\sqrt{\kappa_{i}}\sigma_{i}\right]\rho_{J}dt\nonumber \\
 & +\left(dW_{Q}\mathcal{H}\left[\sqrt{\kappa_{c}}c\right]+dW_{P}\mathcal{H}\left[i\sqrt{\kappa_{c}}c\right]\right)\rho_{J}\nonumber \\
 & +\underset{i}{\sum}\left(dW_{x,i}\mathcal{H}\left[\sqrt{\kappa_{i}}\sigma_{i}\right]+dW_{y,i}\mathcal{H}\left[i\sqrt{\kappa_{i}}\sigma_{i}\right]\right)\rho_{J},\nonumber 
\end{align}
where the deterministic component is governed by the Lindblad superoperator $\mathcal{D}$ defined as 
\begin{equation}
\mathcal{D}\left[a\right]\rho_{J}=a\rho_{J} a^{\dagger}-\frac{1}{2}\left(a^{\dagger}a\rho_{J}+\rho_{J} a^{\dagger}a\right),\label{eq:Lindblad}
\end{equation}
while the stochastic component is governed by the superoperator $\mathcal{H}$ defined as
\begin{equation}
\mathcal{H}\left[a\right]\rho_{J}=a\rho_{J}+\rho_{J}a^{\dagger}-\left\langle a+a^{\dagger}\right\rangle _{J}\rho_{J},\label{eq:Stochastic}
\end{equation}
for any stochastic collapse operator $a$. The continuous quantum measurements of the observables $Q$, $P$, $\sigma_{x,i}$, and $\sigma_{y,i}$ are associated with the corresponding stochastic collapse operators $\sqrt{\kappa_{c}}c$, $i\sqrt{\kappa_{c}}c$, $\sqrt{\kappa_{i}}\sigma_{i}$, and $i\sqrt{\kappa_{i}}\sigma_{i}$, respectively \cite{Wiseman2010,BvHJ07,Nurdin14}. The inherent randomness in the measurement outcomes is related to the Wiener increments, $dW_{Q\left(P\right)}$ and $dW_{x\left(y\right),i}$, each of which is drawn from a Gaussian distribution with a standard deviation of $\sqrt{dt}$. The detection efficiencies of the various channels are incorporated into the Wiener increments. Each measurement detects the continuous currents in cavity and atom channels with noises, with the measurement records given by $\left\langle Q\left(P\right)\right\rangle _{J}+dW_{Q\left(P\right)}/dt$ and $\left\langle \sigma_{x\left(y\right),i}\right\rangle _{J}+dW_{x\left(y\right),i}/dt$, where the expectation values are computed using $\rho_{J}$ from Eq.~(\ref{eq:sme}). The actual features used for training and testing QRC, denoted as $x_{n}(t)$, average measurement records over multiple measurement trajectories.

In the ideal scenario where the number of measurements approaches infinity, the impacts of the measurement back-actions, $dW_{Q\left(P\right)}$ and $dW_{x\left(y\right),i}$, are averaged out. This idealization is described by the deterministic master equation 
\begin{equation}
\frac{d\rho}{dt}=-i\left[H_{0}+H_{1}\left(t\right),\rho\right]+2\mathcal{D}\left[\sqrt{\kappa_{c}}c\right]\rho+2\underset{i}{\sum}\mathcal{D}\left[\sqrt{\kappa_{i}}\sigma_{i}\right]\rho,\label{eq:me}
\end{equation}
accompanied by the averaged measurement records $\left\langle Q\left(P\right)\right\rangle $ and $\left\langle \sigma_{x\left(y\right),i}\right\rangle $, where the expectation values are computed using the density operator $\rho$ from Eq.~(\ref{eq:me}). To roughly maintain a constant total decay rate $\kappa$ as $N_{atom}$ increases, the decay rate of the cavity or each atom is assumed to be $\kappa_{c\left(i\right)}=\kappa/\left(2N_{atom}+2\right)$. A higher $\kappa$ reduces the uncertainty in the feature measurement, commonly referred to a strong measurement, while a lower $\kappa$ corresponds to a weak measurement \cite{Fuchs2001}. 

To establish a linear regression for training and testing the reservoir computer, the expectation values of observables are linked to a feature vector. For single-atom QRC, the feature vector is explicitly written as 
\begin{equation}
\mathbf{x}(t)=\left[\left\langle Q\right\rangle ,\left\langle P\right\rangle ,\left\langle \sigma_{x1}\right\rangle ,\left\langle \sigma_{y1}\right\rangle \right].\label{eq:feature}
\end{equation}
For multi-atom QRC, all observable expectations are included in the feature vector, leading to $N_{feature}=2N_{atom}+2$, with $N_{feature}$ and $N_{atom}$ the number of features and atoms, respectively. To enhance performance, a polynomial regression is also employed by incorporating both the linear and all the quadratic terms of these expectation values into the feature vector: 
\begin{align}
\mathbf{x}(t) & =[\left\langle Q\right\rangle ,\left\langle P\right\rangle ,\left\langle \sigma_{x1}\right\rangle ,\left\langle \sigma_{y1}\right\rangle ,\label{eq:feature_polynomial}\\
 & \left\langle Q\right\rangle ^{2},\left\langle P\right\rangle ^{2},\left\langle \sigma_{x1}\right\rangle ^{2},\left\langle \sigma_{y1}\right\rangle ^{2},\nonumber \\
 & \left\langle Q\right\rangle \left\langle P\right\rangle ,\left\langle Q\right\rangle \left\langle \sigma_{x1}\right\rangle ,\left\langle Q\right\rangle \left\langle \sigma_{y1}\right\rangle ,\nonumber \\
 & \left\langle P\right\rangle \left\langle \sigma_{x1}\right\rangle ,\left\langle P\right\rangle \left\langle \sigma_{y1}\right\rangle ,\left\langle \sigma_{x1}\right\rangle \left\langle \sigma_{y1}\right\rangle ].\nonumber 
\end{align}
For multi-atom QRC, this results in $N_{feature}=2N_{atom}^{2}+7N_{atom}+5$ for polynomial regression.

Quantum reservoir computing is commonly set in discrete-time \cite{Fujii2017}. The discretized time is expressed as $t_{k}=k\Delta t$, where the integer $k$ denotes the time index and $\Delta t$ represents the time step. During the time interval from $t_{k-1}$ to $t_{k}$, the discretized input maintains a constant value $f_{k}=f\left(t_{k}\right)$, where $f\left(t_{k}\right)$ originates from Eq.~(\ref{eq:H1}). The discretized features are sampled at $t_{k}$ as $x_{kn}=x_{n}\left(t_{k}\right)$, where a constant bias term $x_{k0}=1$ is also introduced. 

The relationship between the feature vector and output is approximately inferred through a learning process. Let $\bar{y}_{k}$ represent the target output capturing some key features of the input $f_{k}$. The objective of training is to determine the weights $W_{n}$ in order to achieve the actual output
\begin{equation}
y_{k}=\sum_{n=1}^{N_{feature}}x_{kn}W_{n},\label{eq:yk}
\end{equation}
such that the normalized root mean square error 
\begin{equation}
{\rm NRMSE}=\frac{1}{\bar{y}_{max}-\bar{y}_{min}}\sqrt{\frac{\sum_{k=1}^{L}\left(y_{k}-\bar{y}_{k}\right)^{2}}{L}}\label{eq:NRMSE}
\end{equation}
is minimized, where $L$ is the number of time steps in the training period, and $\bar{y}_{max}$ and $\bar{y}_{min}$ are the maximum and minimum of the target time series output, respectively. The approach for this minimization is discussed in App.~\ref{sec:NRMSEtraining}. To test the performance of reservoir computing, newly generated $x_{kn}$ and $\bar{y}_{k}$ in the testing period, along with the trained weights $W_{n}$, are utilized to calculate the NRMSE. 

\begin{figure}
\includegraphics[width=1.0\linewidth]{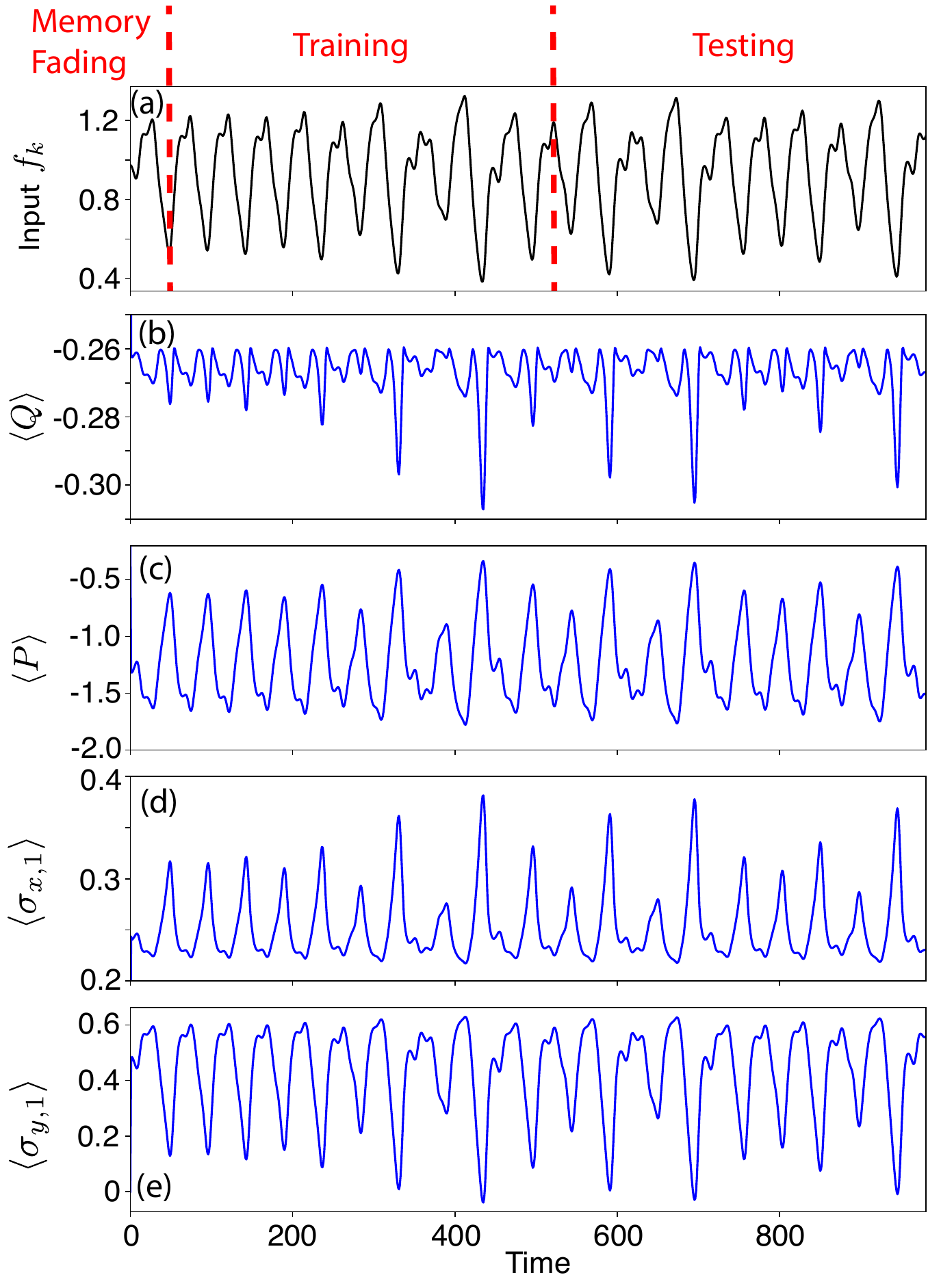}
\caption{Input and features for the Mackey-Glass task. (a) The input function, $f_{k}$, divided into memory fading, training, and testing zones. (b)-(e) The corresponding feature from a single-atom reservoir with $\omega_{1}=20$ and $g_{1}=30$. Parameters: $\tau=20$, $\kappa=10$, $\omega_{c}=40$, and $\epsilon=20$.}
\label{Mackey_Glass_Input_Readouts}
\end{figure}

\section{QRC Performance Analysis}

\subsection{Mackey-Glass task}

As a benchmark for long-term memory, the Mackey-Glass task requires the reservoir to retain past information from the input function in order to predict its future behavior. The input function is governed by the Mackey-Glass equation
\begin{equation}
\frac{df\left(t\right)}{dt}=\frac{\beta f\left(t-\tau_{M}\right)}{1+f^{10}\left(t-\tau_{M}\right)}-\gamma f\left(t\right),\label{eq:MGEq}
\end{equation}
where the parameters $\beta=0.2$, $\gamma=0.1$, and $\tau_{M}=17$ are standard values typically used in the chaotic regime \cite{Fujii2017,Dudas2023,Hulser2023}. To implement this task, a buffer is introduced by discarding the first $1000$ time units in Eq.~ (\ref{eq:MGEq}), and thereby the discretized function $f_{k}=f\left(kdt+1000\right)$ is considered as the actual input, with discretized time $t=kdt$, as shown in Fig.~\ref{Mackey_Glass_Input_Readouts}(a). The time series is sampled with a time step of $dt=1$. The target output, $\bar{y}_{k}$, defined as $\bar{y}_{k}=f_{k+\tau}$, is intended to forecast the future evolution of the input function with a time delay $\tau$.

The proposed quantum optical reservoir performs well in the Mackey-Glass task, as it exhibits discernible responses to various input waveforms, which are quantified by the features of observables illustrated in Figs.~\ref{Mackey_Glass_Input_Readouts}(b)-(e). The different input waveform shapes in Figs.~\ref{Mackey_Glass_Input_Readouts}(a) correspond to the varying feature shapes observed in Figs.~\ref{Mackey_Glass_Input_Readouts}(b)-(e). The simulation time is partitioned into three distinct intervals: memory fading, training, and testing. The memory fading ensures that, during both training and testing, the features are influenced exclusively by the input function, rather than by the initial state of the quantum reservoir. 

\begin{figure}
\includegraphics[width=1.0\linewidth]{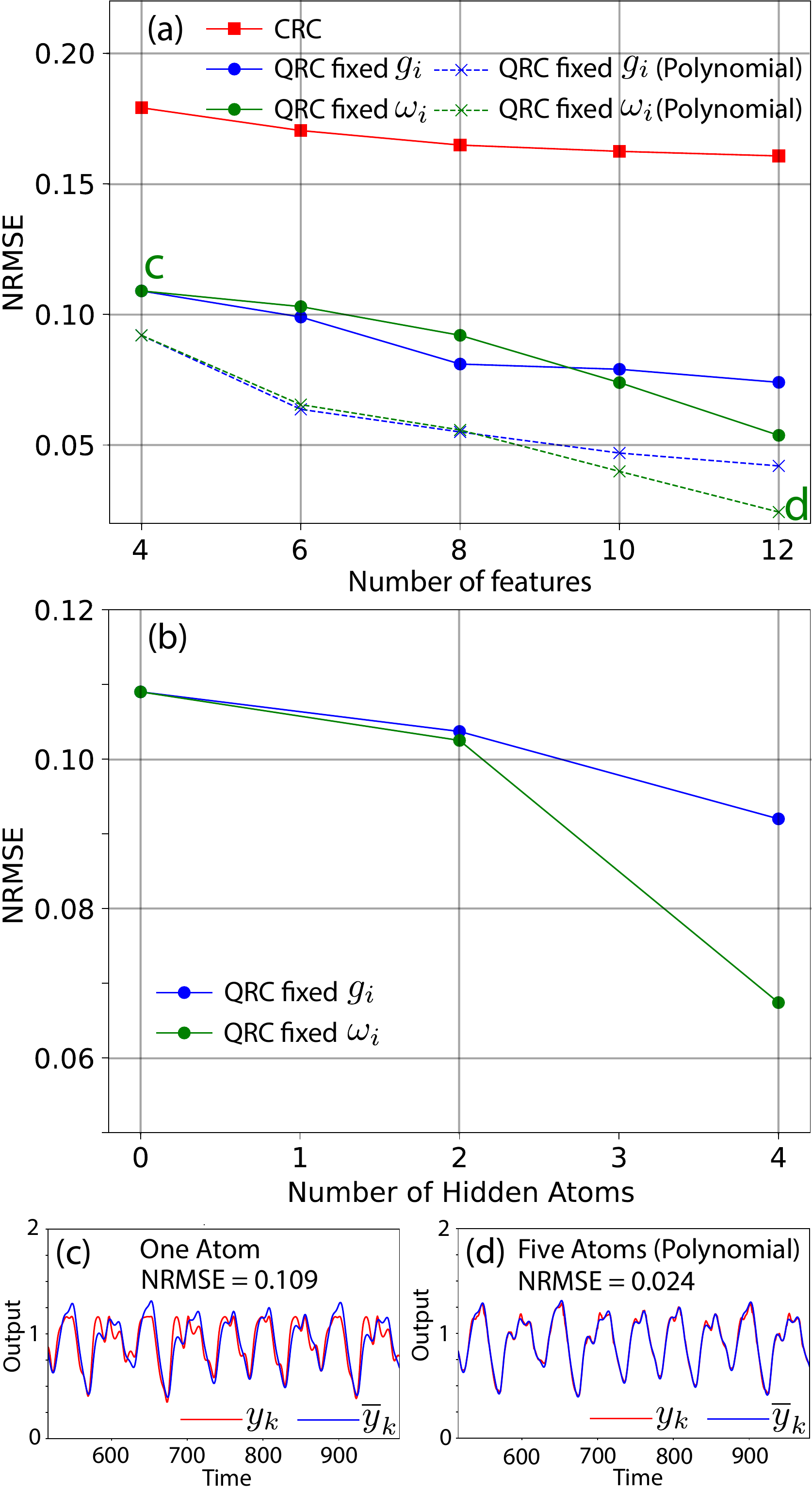}
\caption{Testing result for the Mackey-Glass task with various reservoir scales. (a) NRMSE plotted against the number of features, where features are obtained from all atoms and the photon mode. Blue lines: fixed $g_{i}=30$ for all atoms and $\omega_{i}=20$ for one atom, $\omega_{i}=[0,40]$ for two atoms, $\omega_{i}=[0,20,40]$ for three atoms, $\omega_{i}=[0,10,30,40]$ for four atoms, and $\omega_{i}=[0,10,20,30,40]$ for five atoms. Green lines: fixed $\omega_{i}=20$ for all atoms and $g_{i}=30$ for one atom, $g_{i}=[10,50]$ for two atoms, $g_{i}=[10,30,50]$ for three atoms, $g_{i}=[10,20,40,50]$ for four atoms, and $g_{i}=[10,20,30,40,50]$ for five atoms. Solid lines: regular linear regression. Dashed lines: polynomial regression incorporating both linear and quadratic terms of observables. Red line: CRC averaging $1000$ random trajectories on the echo state network. (b) NRMSE as a function of the number of hidden (unmeasured) atoms in QRC, while maintaining measurement of $4$ specified observables from the cavity field and the particular atom with $g_{1}=30$ and $\omega_{1}=20$. (c)-(d) Actual (red) and target (blue) outputs from QRC with one atom (linear regression) and five atoms (polynomial regression), corresponding to the points marked by letters ``c'' and ``d'' in panel (a), respectively. Parameters: $\tau=20$, $\kappa=10$, $\omega_{c}=40$, $\epsilon=20$. }
\label{Mackey_Glass_Change_AtomNum}
\end{figure}

Figure~\ref{Mackey_Glass_Change_AtomNum} illustrates the performance enhancement as the quantum reservoir scales up, demonstrating the scalability of QRC. The scale of QRC is determined by two factors: (i) the number of observables, which is determined by the number of channels one wishes to measure, with a maximum availability of $2N_{atom}+2$; and (ii) the number of basis states spanning the Hilbert space, which scales as $N_{c}2^{N_{atom}}$, with $N_{c}$ the number of involved photon Fock states. In QRC, the number of observables is also referred to as the number of features. The blue and green solid lines in Fig.~\ref{Mackey_Glass_Change_AtomNum}(a) shows the decrease in NRMSE as the number of atoms increases from $1$ to $5$, where measurements of the cavity field and all atoms lead to a corresponding increase in the number of features from $4$ to $12$. For comparison, the red solid line in Fig.~\ref{Mackey_Glass_Change_AtomNum}(a) shows the performance of CRC as a function of the number of features in an echo state network, with further details discussed in App.~\ref{sec:CRC}. The advantage of QRC over CRC is attributed to the exponentially increasing number of quantum basis states underlying the few measured features. The power of increasing basis states is further demonstrated in Fig.~\ref{Mackey_Glass_Change_AtomNum}(b), where the number of measured features is fixed at $4$ (including $2$ cavity and $2$ atomic observables), while additional hidden (unmeasured) atoms is introduced to expand the computational Hilbert space. The improved performance for a fixed number of measured features with an increase in the dimension of the QRC Hilbert space has previously been observed for Ising models of QRC \cite{Fujii2017,chen2019learning}. A possible  reason for this could be that, for the particular tasks considered, more complex fading memory maps generated by the QRC as the Hilbert space is increased are able to better capture features in the task to be learned. However, the improvement is expected to eventually plateau for a high enough dimension of the Hilbert space.  

The outcome from the polynomial regression, incorporating both linear and quadratic terms of all observables, is depicted by the dashed lines in Fig.~\ref{Mackey_Glass_Change_AtomNum}(a). The result is consistent with previous research indicating that appending nonlinear features from a reservoir can notably enhance performance \cite{Araujo2020,Govia2021}. A comparison between two extreme cases, linear regression with one atom and polynomial regression with five atoms, is illustrated in Figs.~\ref{Mackey_Glass_Change_AtomNum}(c)(d), demonstrating a substantial performance improvement resulting from the combination of scalability and polynomial regression.

\begin{figure}
\includegraphics[width=1.0\linewidth]{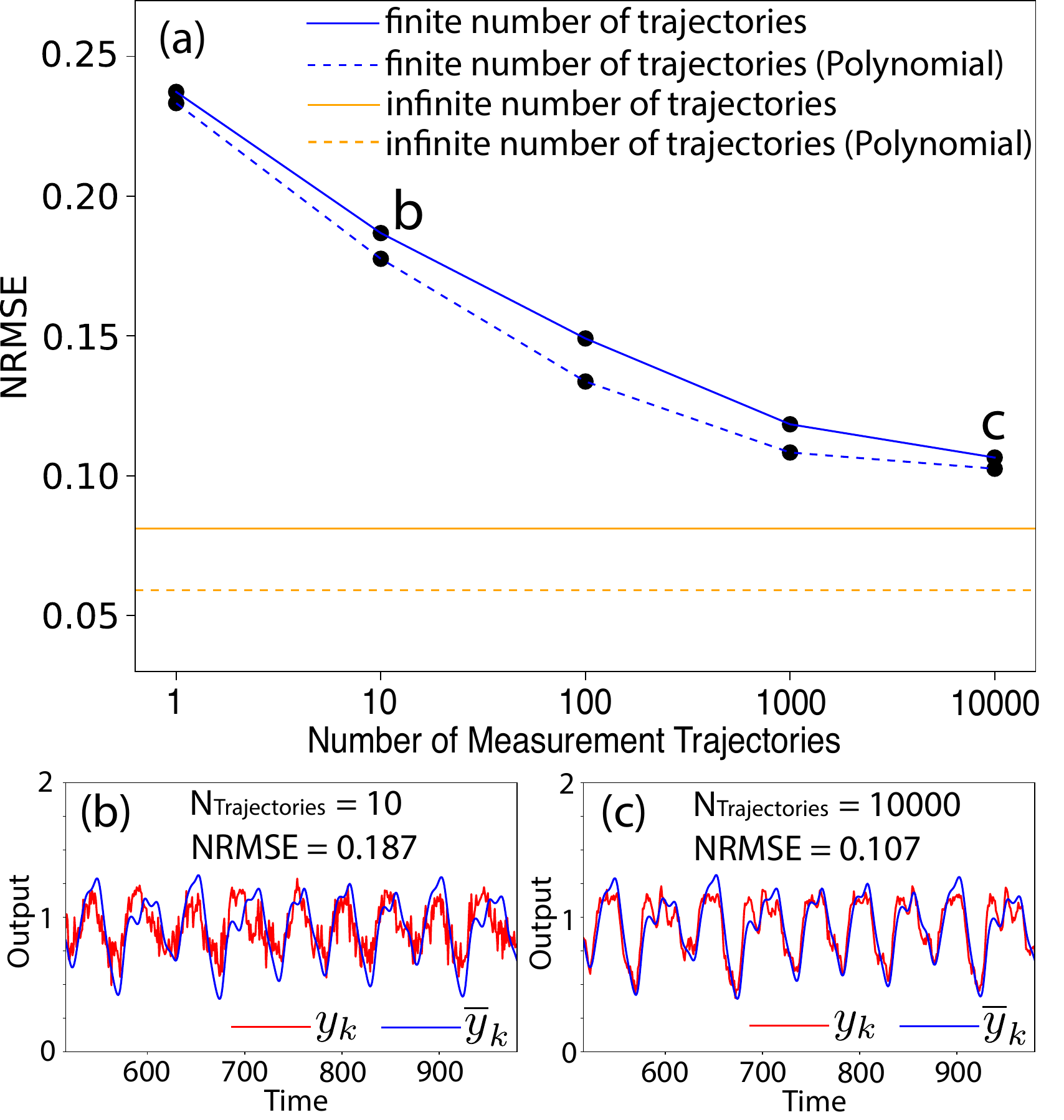}
\caption{Testing results for the Mackey-Glass task within the framework of continuous quantum measurement, where each observable feature is obtained by averaging measurement records over multiple measurement trajectories. (a) NRMSE as a function of the number of measurement trajectories for a three-atom QRC with $g=30$ and $\omega_{i}=[0,20,40]$. Blue lines: averaged results from a finite number of trajectories simulated with stochastic master equation in Eq.~(\ref{eq:sme}). Orange lines: asymptotic results from infinite number of trajectories simulated with deterministic master equation in Eq.~(\ref{eq:me}). Solid lines: regular regression. Dashed lines: polynomial regression. (b)(c) The actual (red) and target (blue) outputs for $10$ and $10000$ trajectories, corresponding to the points labeled "b" and "c" in panel (a), respectively. Parameters: $\tau=20$, $\kappa=10$, $\omega_{c}=40$, $\epsilon=20$, $dt=1$.}
\label{Mackey_Glass_Trajectories}
\end{figure}

While the results presented in Fig.~\ref{Mackey_Glass_Change_AtomNum} are based on observables averaged over an infinite number of trajectories (simulated via the deterministic master equation in Eq.~(\ref{eq:me})), the number of measurements plays a crucial role in determining the QRC performance under realistic conditions \cite{Hu2023, Palacios2024}. A methodology based on eigentasks has been proposed to evaluate the impact of finite sampling noise on the expressive capacity of quantum machine learning models \cite{Hu2023}. The effect of finite number of trajectories in continuous measurement has also been studied for QRC to process quantum inputs \cite{khan2021}. The effect of using a finite number of measurement trajectories is illustrated in Fig.~\ref{Mackey_Glass_Trajectories} for three-atom QRC. To simulate the stochastic nature of measurement noise, the stochastic master equation in Eq.~(\ref{eq:sme}) is numerically evolved multiple times, each with independent realizations of the random Wiener increments $dW_{Q\left(P\right)}$ and $dW_{x\left(y\right),i}$. Each full evolution corresponds to a single measurement trajectory. The observable features, $x_{kn}$, are then obtained by averaging the measurement records across all such trajectories. As shown in Fig.~\ref{Mackey_Glass_Trajectories}(a), increasing the number of trajectories leads to improved QRC performance, with results asymptotically approaching those obtained from an infinite number of trajectories.  Notably, the performance enhancement achieved by polynomial regression is reduced under finite trajectory conditions (i.e. the performance gap between the two blue curves is smaller than that between the two orange curves). This implies that the polynomial regression would require many more samples to resolve the optimal weights as the number of weights increased compared to the linear regression case.

\begin{figure}
\includegraphics[width=1.0\linewidth]{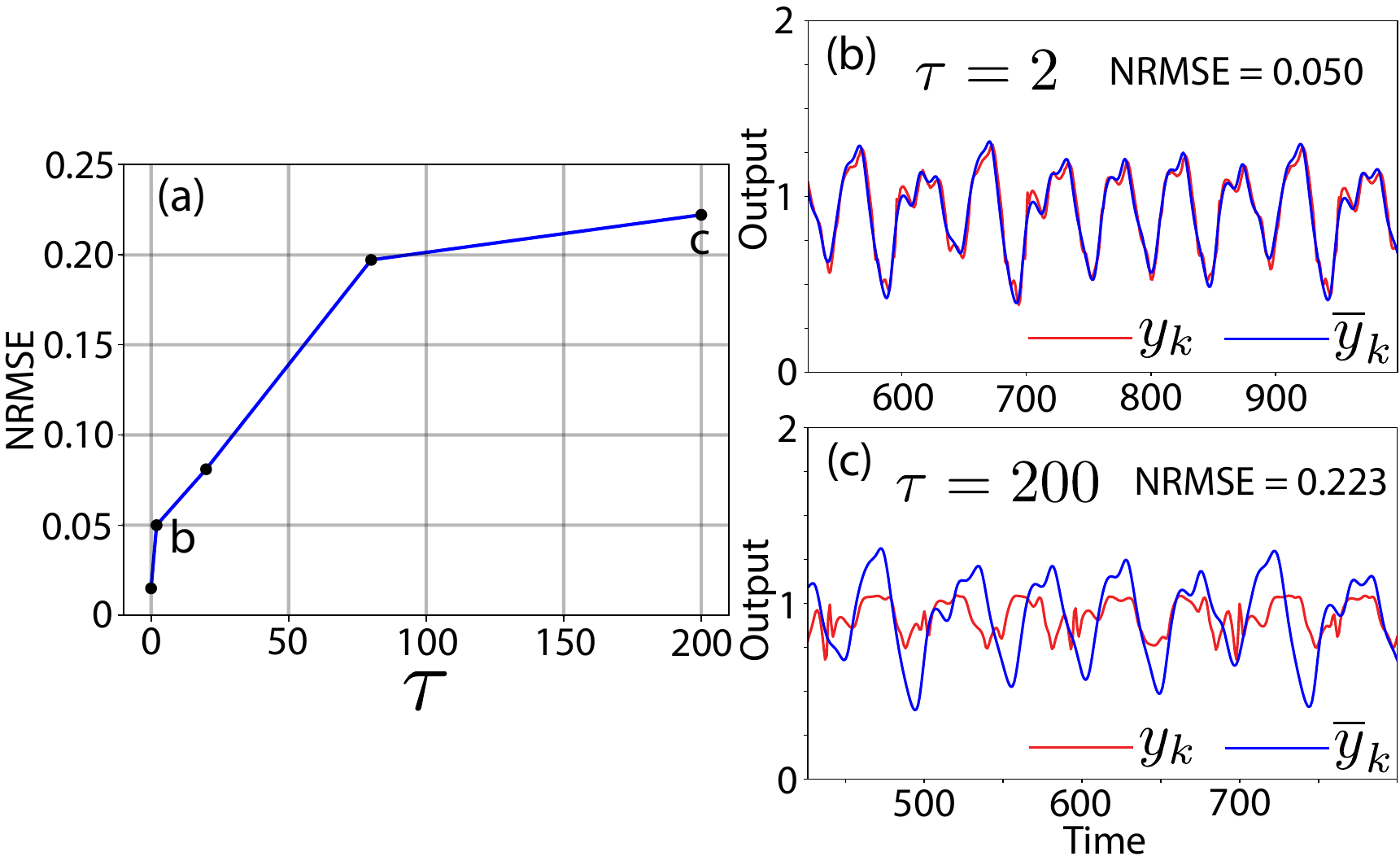}
\caption{Testing result for the Mackey-Glass task with various delay $\tau$. (a) NRMSE as a function of delay $\tau$ for a three-atom QRC with $g=30$ and $\omega_{i}=[0,20,40]$. (b)(c) The actual (red) and target (blue) outputs with $\tau=2$ and $\tau=200$, corresponding to the points marked by letters ``b'' and ``c'' in panel (a), respectively. Parameters: $\kappa=10$, $\omega_{c}=40$, $\epsilon=20$.}
\label{Mackey_Glass_Change_Delay}
\end{figure}

\begin{figure}
\includegraphics[width=1.0\linewidth]{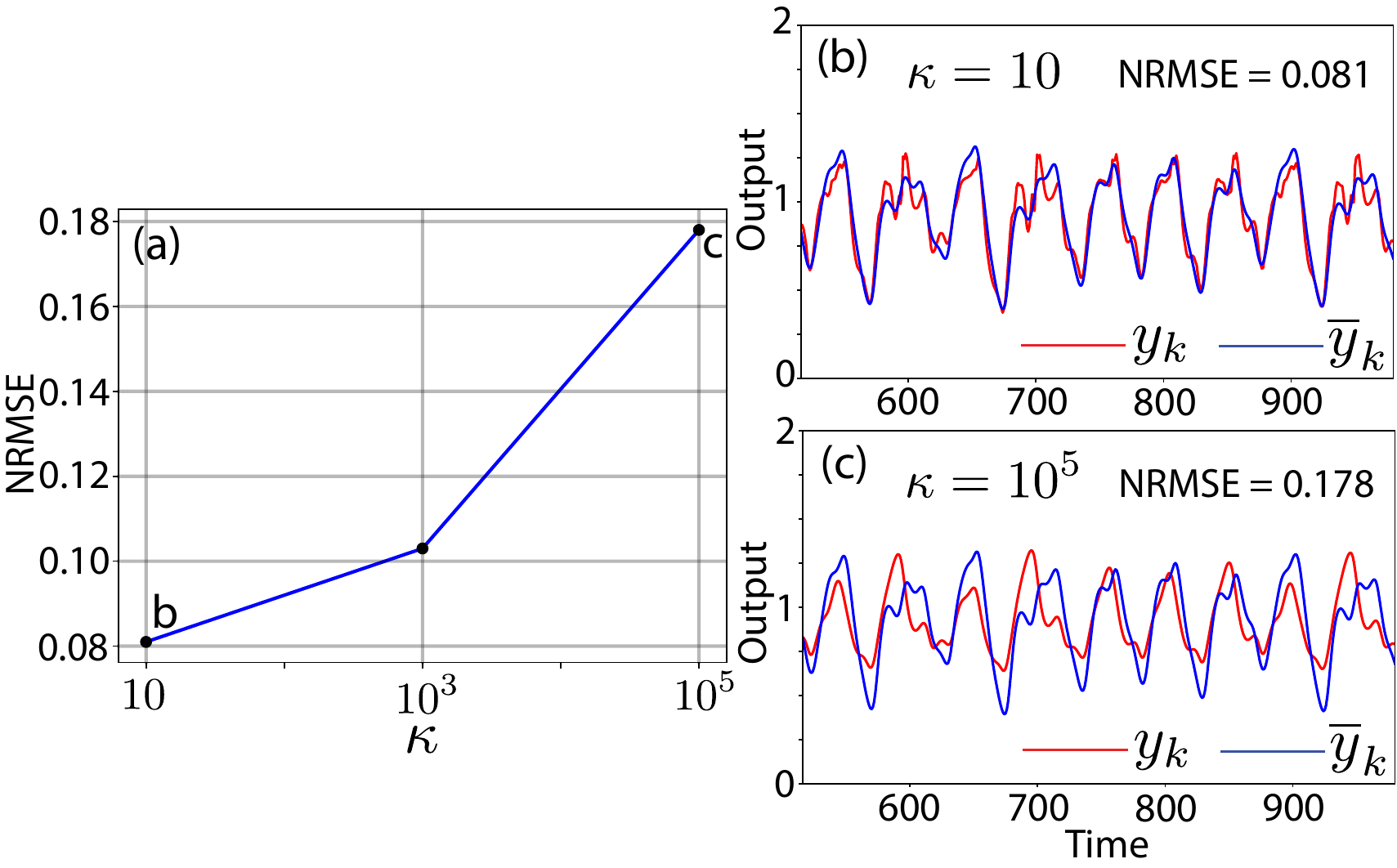}
\caption{Testing result for the Mackey-Glass task with various decay rate $\kappa$. (a) NRMSE as a function of $\kappa$ for three-atom QRC with $g=30$ and $\omega_{i}=[0,20,40]$. (b)(c) The actual (red) and target (blue) outputs with $\kappa=10$ and $\kappa=10^{5}$, corresponding to the points marked by letters ``b'' and ``c'' in panel (a), respectively. Parameters: $\tau=20$, $\omega_{c}=40$, $\epsilon=20$. }
\label{Mackey_Glass_Change_Kappa}
\end{figure}

The effects of delay $\tau$ and the decay rate $\kappa$ are illustrated in Figs.~\ref{Mackey_Glass_Change_Delay} and \ref{Mackey_Glass_Change_Kappa}, respectively. According to Eq.~(\ref{eq:MGEq}), an increased $\tau$ requires the reservoir to exhibit stronger memory in order to retain more past information from the input, thereby improving its ability to predict future output, which makes the task more demanding. In contrast, a larger $\kappa$ accelerates the rate at which the reservoir forgets input information due to photon leakage and atomic spontaneous emission.

\begin{figure}
\includegraphics[width=1.0\linewidth]{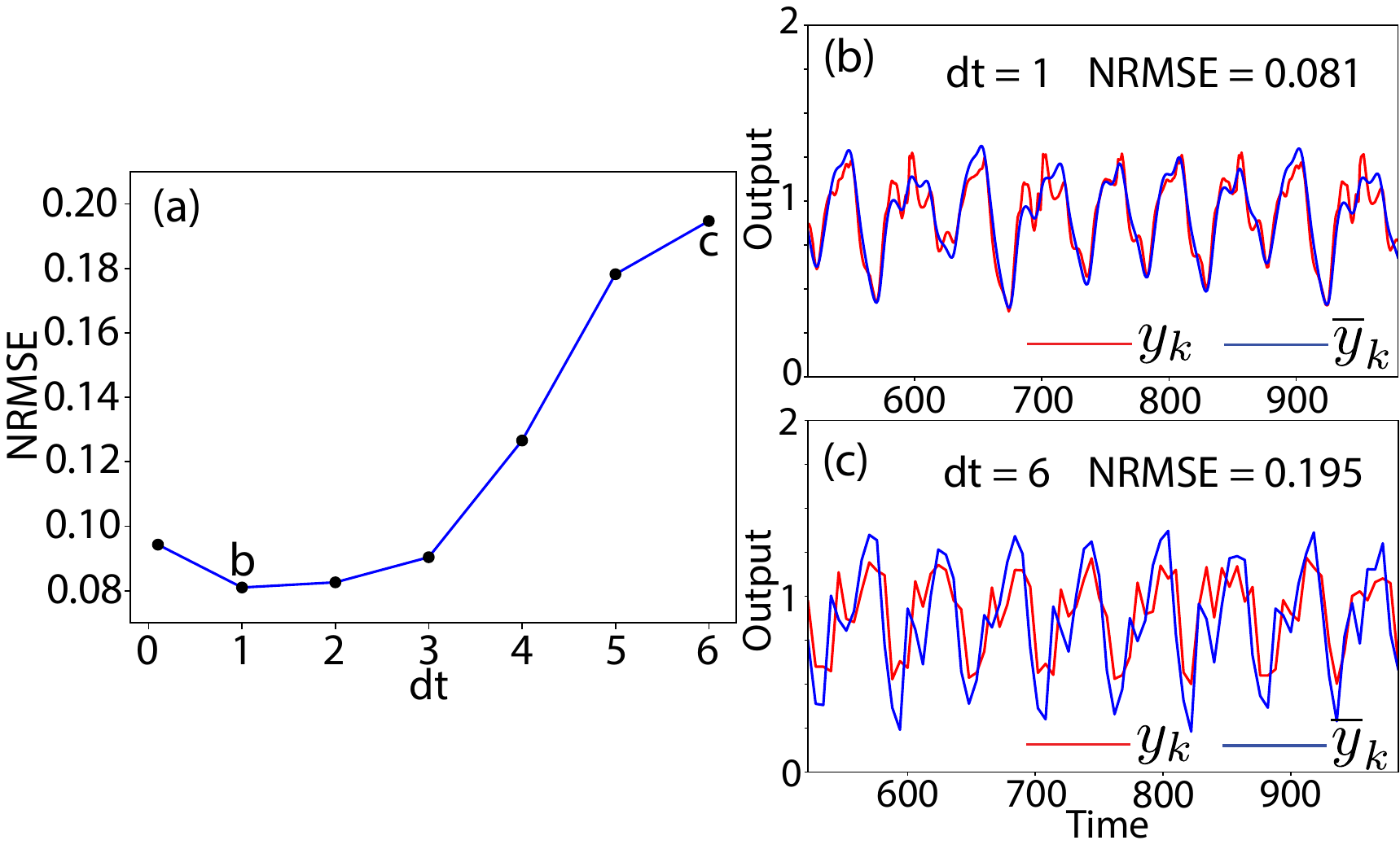}
\caption{Testing result for the Mackey-Glass task with various $dt$ in Eq.~(\ref{eq:MGEq}). (a) NRMSE as a function of $dt$ for a three-atom QRC with $g=30$ and $\omega_{i}=[0,20,40]$. (b)(c) The actual (red) and target (blue) outputs with $dt=1$ and $dt=6$, corresponding to the points labeled ``b'' and ``c'' in panel (a), respectively. Parameters: $\tau=20$, $\kappa=10$, $\omega_{c}=40$, $\epsilon=20$.}
\label{Mackey_Glass_dt}
\end{figure}

The results presented above for the Mackey-Glass task were obtained using a time step of $dt=1$ in Eq.~(\ref{eq:MGEq}). However, varying the value of $dt$ yields different variants of the Mackey-Glass task. It is important to note that the time step used in the numerical simulation must be significantly smaller than the $dt$ parameter specified here. To ensure numerical convergence in simulating the quantum evolution, our master equation simulator performs $4000$ substeps between $t$ and $t+dt$. As illustrated in Fig.~\ref{Mackey_Glass_dt}(a), the QRC scheme exhibits relatively stable performance for tasks with $dt \le 3$. For larger values of $dt$, the NRMSE tends to increase not only because the density of the training and testing datasets is reduced, but also because the samples generated by numerically solving the Mackey-Glass equation using a first-order method with a larger step size $dt$ deviate more from the true solution. This effect is evident in the non-smooth target output (blue line) shown in Fig.~\ref{Mackey_Glass_dt}(c).

\begin{figure}
\includegraphics[width=1.0\linewidth]{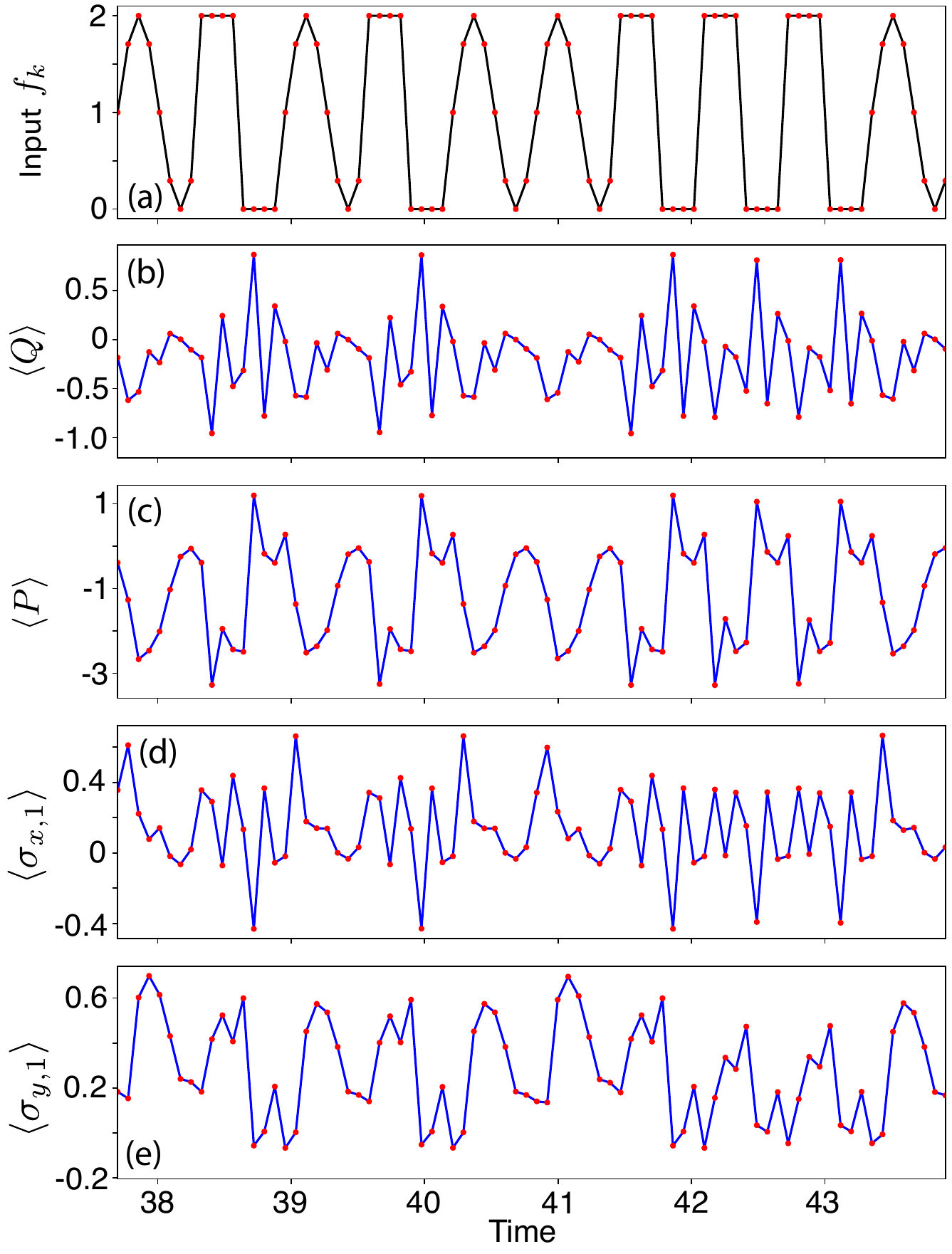}
\caption{Input and features for the sine-square waveform classification task, where $110$ random waveforms are sent in as the input, with $10$ waveforms allocated for memory fading, $50$ waveforms for training, and $50$ waveforms for testing. (a) A segment of the input function, $f_{k}$, representing the first $10$ waveforms during the testing zone. (b)-(e) The corresponding features from a single-atom reservoir with $\omega_{1}=20$ and $g_{1}=30$. Parameters: $\omega_{c}=40$, $\kappa=10$, $\epsilon=20$, $\omega_{ss}=10$, $N_{ss}=8$.}
\label{Sine_Square_Input_Readouts}
\end{figure}

\subsection{Sine-square waveform classification task}

The purpose of the sine-square waveform classification task is to determine whether each input data point corresponds to a sine or square waveform. The time-dependent input, denoted as $f_k$, consists of 110 randomly generated sine and square waveforms. Of these, 10 are reserved for memory fading, 50 for training, and 50 for testing. Each waveform is discretized into $N_{ss}$ points, yielding a time step $\Delta t = 2\pi / \left( N_{ss} \omega_{ss} \right)$, where $\omega_{ss}$ represents the oscillation frequency of the input. Figure~\ref{Sine_Square_Input_Readouts}(a) illustrates these discretized time points (denoted as red dots) for the first 10 waveforms during the testing phase. The target output, $\bar{y}_k$, which is used to classify the input signal, is set to 0 if the input belongs to a square waveform and 1 if it corresponds to a sine waveform.

The sine-square waveform classification represents a nonlinear task, requiring the reservoir to process a linearly inseparable input dataset that exhibits abrupt transitions. In a linear, closed quantum system, the smoothly varying expectation values of the features, $\left\langle c\right\rangle \propto\exp\left(-i\omega_{c}t\right)$ and $\left\langle \sigma_{i}\right\rangle \propto\exp\left(-i\omega_{i}t\right)$, fail to capture the high-frequency, sudden changes present in the input. The incorporation of nonlinearity, arising from the pumping and decay processes in an open quantum system, enhances the reservoir's ability to rapidly respond to these abrupt changes. This is clearly demonstrated by the distinct feature measurements shown in Figs.~\ref{Sine_Square_Input_Readouts}(b)-(e), which directly correspond to the input waveforms depicted in Fig.~\ref{Sine_Square_Input_Readouts}(a). The sine and square input waveforms in Figs.~\ref{Sine_Square_Input_Readouts}(a) correspond to the different feature shapes observed in Figs.~\ref{Sine_Square_Input_Readouts}(b)-(e).

\begin{figure}
\includegraphics[width=1.0\linewidth]{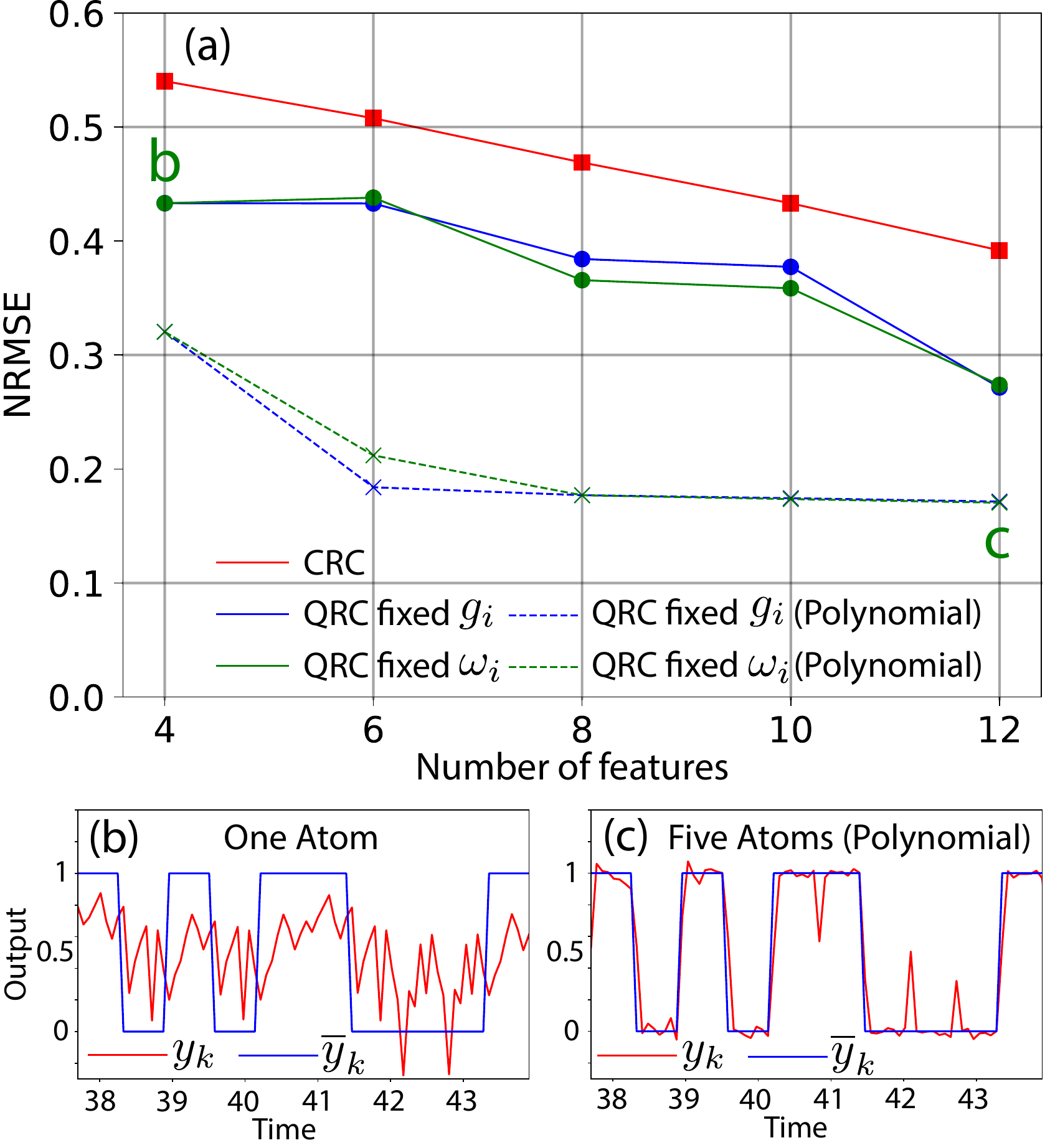}
\caption{Testing result for the sine-square waveform classification task with various reservoir scales. (a) NRMSE plotted against the number of features, featuring results from QRC with linear regression, QRC with polynomial regression, and CRC on echo state network, where the parameters $\omega_{i}$ and $g_{i}$ align with those in Fig.~\ref{Mackey_Glass_Change_AtomNum}. (b)(c) The actual (red) and target (blue) outputs from QRC with one atom (linear regression) and five atoms (polynomial regression), corresponding to the points marked by letters \textquotedblleft b\textquotedblright{} and \textquotedblleft c\textquotedblright{} in panel (a), respectively. Parameters: $\omega_{c}=40$, $\kappa=10$, $\epsilon=20$, $\omega_{ss}=10$, and $N_{ss}=8$.}
\label{Sine_Square_Change_AtomNum}
\end{figure}

\begin{figure}
\includegraphics[width=1.0\linewidth]{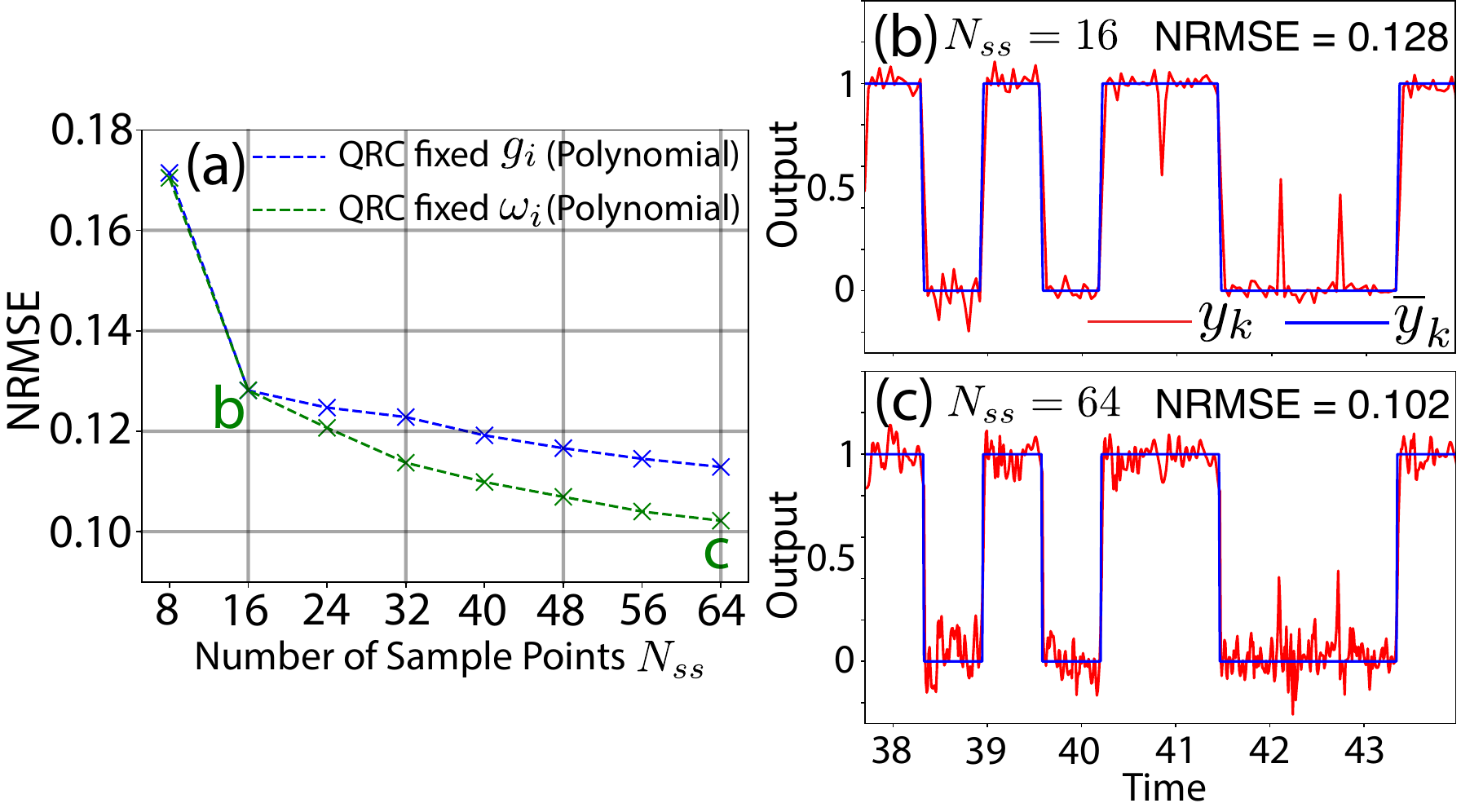}
\caption{Testing result for the sine-square waveform classification task with various numbers of sample points, $N_{ss}$, within each period of the sine or square waveform. (a) NRMSE as a function of $N_{ss}$ for two sets of five-atom QRCs with $g_{i}=30$ and $\omega_{i}=[0,10,20,30,40]$ (blue) and with $\omega_{i}=20$ and $g_{i}=[10,20,30,40,50]$ (green), utilizing polynomial regression. The point at $N_{ss}=8$ corresponds to the \textquotedbl c\textquotedbl{} marker in Fig.~\ref{Sine_Square_Change_AtomNum}(a). (b)(c) The actual (red) and target (blue) outputs with $N_{ss}=16$ and $N_{ss}=64$, corresponding to the \textquotedbl b\textquotedbl{} and \textquotedbl c\textquotedbl{} markers in panel (a), respectively. Parameters: $\omega_{c}=40$, $\kappa=10$, $\epsilon=20$, and $\omega_{ss}=10$.}
\label{Sine_Square_SamplePoints}
\end{figure}

Figure.~\ref{Sine_Square_Change_AtomNum} illustrates the performance improvement in account of scalability and polynomial regression, along with the comparison between QRC and CRC. It is observed that the performance associated with polynomial regression, represented by the blue and green dashed lines in Fig.~\ref{Sine_Square_Change_AtomNum}(a), tends to saturate at $8$ features. This performance saturation suggests that the maximum performance achievable with the current sample size, associated with $N_{ss}=8$, has been attained. The corresponding comparison between the actual and target outputs is illustrated in Fig.~\ref{Sine_Square_Change_AtomNum}(c), where discrepancies primarily occur at the first sample point in each waveform. Figure~\ref{Sine_Square_SamplePoints} shows that these discrepancies can be mitigated by increasing the number of sample points, $N_{ss}$, within each period of the sine or square waveform. This is evidenced by the decrease in NRMSE, as shown in Fig.~\ref{Sine_Square_SamplePoints}(a). The comparison between Fig.~\ref{Sine_Square_SamplePoints}(b) and Fig.~\ref{Sine_Square_SamplePoints}(c) reveals that larger $N_{ss}$ reduces the impact of abrupt shifts in the input waveform on the output. Meanwhile, it also introduces more small oscillations resulting from higher-frequency changes in the input function due to smaller $\Delta t$. 

\section{Discussion}

We have presented a minimalistic QRC platform consisting of up to five atoms with excellent capabilities in quantum memory and nonlinear data processing. The inputs are classical functions integrated into the coefficient of coherent driving, allowing for direct comparison with the performance of classical reservoir computing. Notably, quantum decoherence of the reservoir facilitates the memory fading without the need for external erasing. This means that features are solely determined by the input, rather than by the initial state, after a certain period of quantum dissipation. Compared to other schemes of quantum reservoir computing, our paradigm has two major advantages: practicality and scalability. 

In terms of practicality, continuous quantum measurement utilizing the homodyne detections of cavity quadratures and atomic spins is considered. These observable detections can be in fact carried out simultaneously \cite{Wei2008,Ruskov2010,Hacohen2016,Ochoa2018} and do not require tomography, making them more feasible than measurements of probability distributions in quantum basis states proposed in previous works \cite{Fujii2017,Ghosh2019,Dudas2023,Angelatos2021}. 

The quantum optical reservoir we present offers enhanced scalability when compared to reservoirs based on quantum networks, such as the Ising \cite{Fujii2017} and Fermi-Hubbard \cite{Ghosh2019} models. This advantage stems from the unique coupling mechanism between the atoms and the cavity field. Specifically, the addition of a new atom requires only coupling it to the cavity field, which consequently induces its interaction with the remaining atoms in the reservoir. Moreover, the number of quantum basis states in our reservoir scales proportionally to $2^{N_{atom}}$, enabling faster growth compared to the reservoirs not based on quantum networks, such as a single Kerr nonlinear oscillator \cite{Govia2021} and two coupled linear oscillators \cite{Dudas2023}. The increase in basis states, by itself, is able to improve the performance of QRC, as demonstrated by Fig.~\ref{Mackey_Glass_Change_AtomNum}(b). This exponential scaling is also a crucial factor in the advantage demonstrated by the comparison between QRC and CRC, as shown in Figs.~\ref{Mackey_Glass_Change_AtomNum}(a) and \ref{Sine_Square_Change_AtomNum}(a).

Additionally, previous studies have indicated that appending nonlinear post-processing to the features can enhance the performance of both QRC \cite{Govia2021} and CRC \cite{Paquot2012,Appeltant2011,Brunner2013}. In our work, this idea is implemented through polynomial regression, which additionally appends quadratic combinations of observable expectations to the feature matrix. The results demonstrate a significant improvement in performance. 

We also mention a subsequent contribution \cite{ZENS24} in which the scheme presented in this paper is augmented with the addition of a linear feedback of the features into the input to enhance the expressivity of the system, adopting a proposal in \cite{ENS25} for classical echo state networks. It is shown that the addition of such feedback can indeed offer an enhancement of the capability of the model herein but at the price of increased complexity of the setup, increased computation for optimizing the feedback weights, and an increased number of measurements runs to estimate the expectation values of observable at each time step. The results presented in this article demonstrate surprisingly excellent performance on the tested tasks, even without employing the the  more complex feedback scheme described in \cite{ZENS24}. This finding underscores the inherent merit of this minimalistic hardware.

In our QRC framework, it is essential to vary either the detuning $\omega_{i}$, the coupling strength $g_{i}$, or both, in order to induce non-identical memory states in the atoms. We observed a decline in QRC performance when both parameters were fixed. A promising direction for future research would be to investigate the internal mechanisms of QRC to better understand why the heterogeneity of subsystem capabilities is crucial for maintaining the overall performance of the system.

\section*{Acknowledgements}
We gratefully acknowledge the startup funding provided by Wyant College of Optical Sciences, University of Arizona, which supported the initial development of this research project.

\section*{Data Availability}
The data that support the findings of this article are not publicly available. The data are available from the authors upon reasonable request.

\appendix

\section{Minimization of NRMSE for training}
\label{sec:NRMSEtraining}

The goal of the training process is to determine the optimized weights $W_{n}$ that minimize the normalized root mean square error (NRMSE) as defined in Eq.~(\ref{eq:NRMSE}). Since ${\rm NRMSE^{2}}$, being a quadratic function of $W_{n}$, possesses a global minimum and no local minima, the minimization can be effectively achieved using a pseudoinverse method. The feature values, $x_{kn}$, along with the constant bias term $x_{k0}=1$, are arranged into an $L \times (N_{feature} + 1)$ matrix ${\bf X}$, where $L$ represents the number of time steps used in the training process. The target output, $\bar{y}_{k}$, is organized into an $L \times 1$ column vector ${\bf \bar{Y}}$. The weights, $W_{n}$, are represented in a $(N_{feature} + 1) \times 1$ column vector ${\bf W}$. Consequently, the optimized weight vector ${\bf W}$ that minimizes the NRMSE is determined by \cite{Dion2018,Govia2021}.

\begin{equation}
{\bf W}={\bf X}^{+}{\bf \bar{Y}},\label{eq:W}
\end{equation}
where the Moore-Penrose pseudoinverse 
\begin{equation}
{\bf X}^{+}=\left({\bf X^{{\rm T}}}{\bf X}+\delta{\bf I}\right)^{-1}{\bf X^{{\rm T}}},\label{eq:MoorePenrose}
\end{equation}
where ${\bf I}$ denotes the identity matrix, and $\delta = 10^{-10}$ is a ridge regression parameter employed to mitigate overfitting.

\section{Classical reservoir computing}
\label{sec:CRC}

We use echo state network (ESN) to implement classical reservoir computing: 
\begin{align}
    {\bf x}_{k+1} &= {\rm ReLU}({\bf A x}_k + {\bf B} f_k) \\
    y_k &= {\bf W^{\rm T} x}_k + C,
\end{align}
where ${\bf x}_{k}$ is the vector-valued feature of the ESN with a length of $N_{feature}$, while $f_k$ represents the input. The fitting parameters, ${\bf W}$ and $C$, are obtained through optimization, and $y_k$ represents the output at time step $k$. The activation function employed is the rectified linear unit defined as ${\rm ReLU}(z) = \max(0, z)$ for a scalar $z$. For vector arguments, ${\bf x}$, the function is applied element-wise, such that $({\rm ReLU}({\bf x}))_i = {\rm ReLU}(x_i) = \max(0, x_i)$. An individual ESN is characterized by the selection of the matrix ${\bf A}$ and the vector ${\bf B}$.

In the results presented in Figs.~\ref{Mackey_Glass_Change_AtomNum}(a) and \ref{Sine_Square_Change_AtomNum}(a), the average is computed over $1000$ ESNs, with ${\bf A}$ and ${\bf B}$ chosen randomly. A constraint is applied such that the largest singular value of ${\bf A}$ is less than 1 to ensure convergence. The number of features shown in these figures corresponds to the vector length of ${\bf x}_k$ (and consequently ${\bf W}$), which is determined by $N_{feature}$. The training of ${\bf W}$ and $C$ is performed using linear regression, as described in Section \ref{sec:NRMSEtraining}, where ${\bf W}$ is extended to an $(N_{feature} + 1)$-dimensional vector, with $W_0 = C$.

\section{Lorenz task}

\begin{figure}
\includegraphics[width=1.0\linewidth]{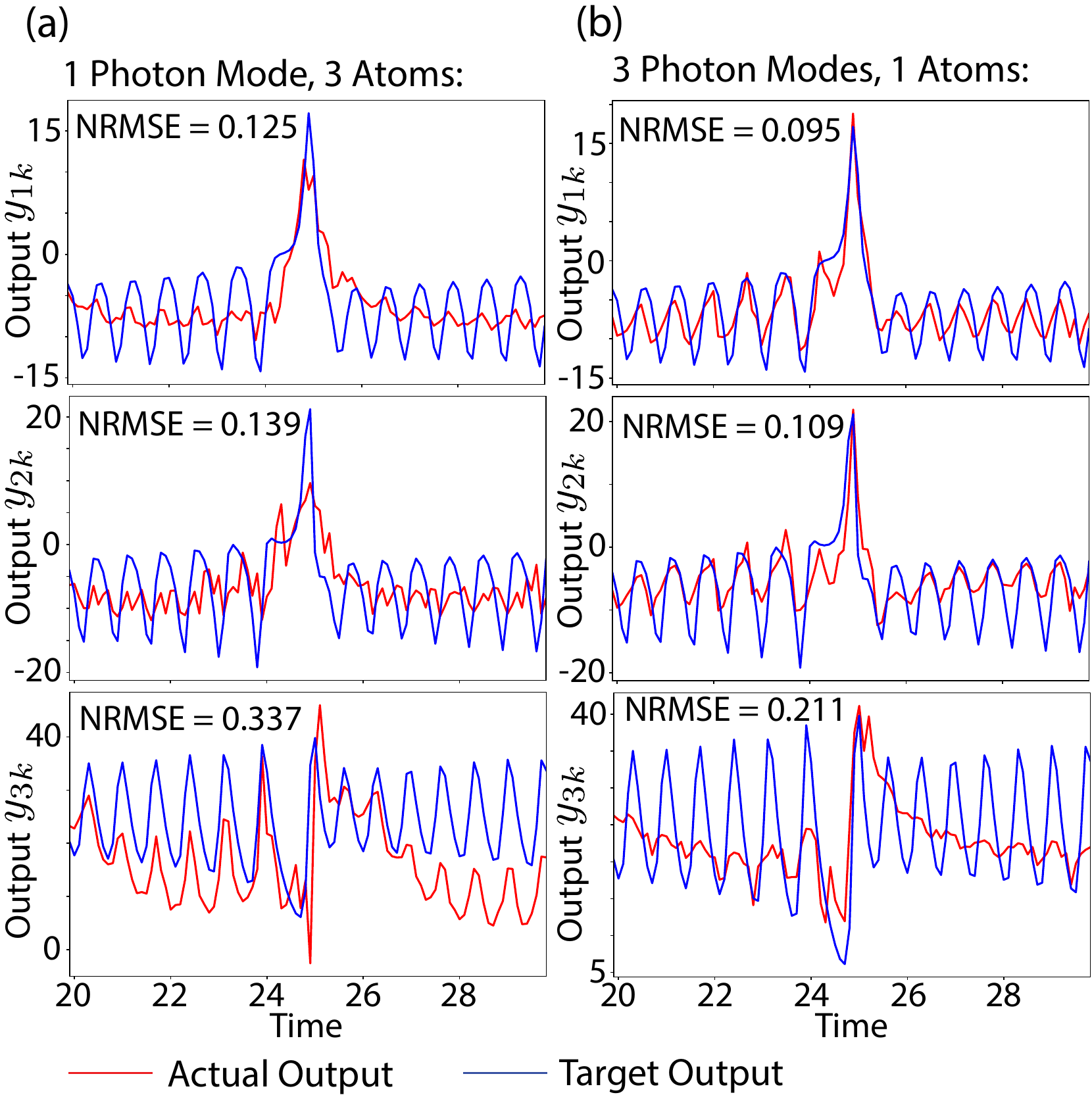}
\caption{Comparison of two different schemes for tackling the Lorenz task. The task aims to forecast the next values of the three Lorenz time series $f_{1,2,3}(t)$ fulfilling Eq.~(\ref{eq:Lorenz}). (a) Outputs from three independent quantum reservoirs, each consisting of a single photon mode with $\omega_{c}=40$ and three atoms with $\omega_{i}=[0,20,40]$. The three input functions are independently encoded into each reservoir via the driving terms $i\epsilon f_{1,2,3}\left(t\right)\left(c-c^{\dagger}\right)$ to the respective single-mode cavities. (b) Outputs from a single quantum reservoir composed of three photon modes with $\omega_{ci}=[0,20,40]$ and one atom with $\omega=40$ in one cavity. The three input functions are encoded simultaneously via the summation of three driving terms $i\epsilon f_{1}\left(t\right)\left(c_{1}-c_{1}^{\dagger}\right) + i\epsilon f_{2}\left(t\right)\left(c_{2}-c_{2}^{\dagger}\right) + i\epsilon f_{3}\left(t\right)\left(c_{3}-c_{3}^{\dagger}\right)$ applied to the three photon modes in this cavity. Red lines: actual outputs. Blue lines: target outputs. The remaining parameters are identical for the two schemes: $\kappa=10$, $\epsilon=20$, $dt=0.1$.}
\label{Lorenz}
\end{figure}

In the main text, we have contemplated two tasks with one dimensional input function. Our atom-cavity QRC can be readily accommodated to address tasks with multidimensional inputs. The Lorenz task is a time series forecasting task with three states, denoted as $f_{1}\left(t\right)$, $f_{2}\left(t\right)$, and $f_{3}\left(t\right)$. The time series is given by the evolution of the Lorenz attractor, fulfilling the equations \cite{Hulser2023, Ahmed2025}
\begin{align}
\frac{df_{1}}{dt} & =\sigma\left(f_{2}-f_{1}\right),\label{eq:Lorenz}\\
\frac{df_{2}}{dt} & =f_{1}\left(\rho-f_{3}\right)-f_{2},\nonumber \\
\frac{df_{3}}{dt} & =f_{1}f_{2}-\beta f_{3},\nonumber 
\end{align}
where $\sigma$, $\rho$, $\beta$ are system parameters and we take $\sigma=10$, $\rho=28$, and $\beta=8/3$ to ensure chaotic behavior of the system. The time series data set is derived numerically via Runge-Kutta method and by taking a time-step size $dt=0.1$. For initial conditions, we use $f_{1}\left(0\right)=0.5$, $f_{2}\left(0\right)=0.1$, $f_{3}\left(0\right)=0.2$ and cut off the first $100$ time units as a transient time before sampling the time series. The task is to predict the next value of these three series, and hence the target outputs are defined as $\overline{y}_{1k}=f_{1}\left(\left(k+1\right)dt\right)$, $\overline{y}_{2k}=f_{2}\left(\left(k+1\right)dt\right)$, and $\overline{y}_{3k}=f_{3}\left(\left(k+1\right)dt\right)$, with the time discretization given by $t=kdt$, and the input $u_k$ at time $k$ is 
$u_k=\left(f_{1}\left(kdt\right),f_{2}\left(kdt\right),f_{3}\left(kdt\right)\right)$. After the transient time period, we evolve the QRC for $30$ time units, with $10$ time units for memory fading, $10$ for training, and $10$ for testing.

We propose and compare two QRC schemes for the Lorenz time series forecasting task:

\textit{Scheme 1: Three single-mode cavities.} In this scheme, the three Lorenz time series are processed independently in three separate quantum reservoirs. Each reservoir comprises a single-mode photon field and three atoms, governed by the Hamiltonian in Eq.~(\ref{eq:H0}) in the main text, with the atom index $i=1,2,3$. Consequently, each reservoir yields $N_{feature}=8$ readout features. The three inputs are independently encoded into each reservoir via the driving term defined in Eq.~(\ref{eq:H1}) of the main text, which is applied to the respective single-mode photon field. The reservoir dynamics are simulated using the master equation in Eq.~(\ref{eq:me}). The three feature vectors and outputs are extracted separately from each reservoir. The comparison between actual and target outputs during the testing period is presented in Fig.~(\ref{Lorenz})(a).

\textit{Scheme 2: One three-mode cavity.} In this scheme, the three Lorenz time series are processed simultaneously in a single quantum reservoir composed of three photon modes and one atom. This configuration also produces $N_{feature}=8$, ensuring a fair comparison with \textit{Scheme 1}. The reservoir Hamiltonian is given by 
\begin{equation}
H_{0}=\underset{i}{\sum}\omega_{ci}c_{i}^{\dagger}c_{i}+\omega\sigma^{\dagger}\sigma+\underset{i}{\sum}g_{i}\left(c_{i}^{\dagger}\sigma+c_{i}\sigma^{\dagger}\right),\label{eq:H0_3modes}
\end{equation}
where $i=1,2,3$ indexes the photon modes, $c_{i}$ represents the annihilation operator for mode $i$, and $\sigma$ denotes the lowering operator of the atom. The three input functions are simultaneously encoded through the combined driving term  
\begin{equation}
H_{1}\left(t\right)=i\epsilon\underset{i}{\sum}f_{i}\left(t\right)\left(c_{i}-c_{i}^{\dagger}\right),\label{eq:H1_3modes}
\end{equation}
which is applied across the three photon modes in the cavity. The system dynamics and measurement process are modeled using the master equation
\begin{equation}
\frac{d\rho}{dt}=-i\left[H_{0}+H_{1}\left(t\right),\rho\right]+2\underset{i}{\sum}\mathcal{D}\left[\sqrt{\kappa_{c}}c_{i}\right]\rho+2\mathcal{D}\left[\sqrt{\kappa_{\sigma}}\sigma\right]\rho,\label{eq:me_3modes}
\end{equation}
where the Lindblad superoperator $\mathcal{D}$ is defined in Eq.~(\ref{eq:Lindblad}) of the main text. The feature vectors and outputs for all three time series are extracted simultaneously from this single reservoir. The comparison between actual and target outputs during the testing period is plotted in Fig.~(\ref{Lorenz})(b).

The performance comparison illustrated in Figs.~\ref{Lorenz}(a) and \ref{Lorenz}(b) demonstrates that the three-mode cavity outperforms the configuration of three independent single-mode cavities, under otherwise identical conditions. This result highlights the potential advantage of leveraging inherent correlations among the input time series through simultaneously evolving multiple cavity fields in QRC, which have higher dimensions (boson fields) than the two-dimensional atoms.

\bibliography{Paper_QRC_No_Feedback}

\end{document}